\documentclass[10pt]{iopart}   
\usepackage{cite}
\pdfoutput=1

\usepackage{graphicx}
\usepackage{color} 
\usepackage{multirow}
\usepackage{bm}
\usepackage{bbold}
\usepackage{xcolor}
\usepackage{epsfig}
\usepackage{amsfonts}

\usepackage{gensymb}

\newcommand{\beqa}{\begin{eqnarray}}
\newcommand{\beeq}{\begin{equation}}
\newcommand{\eeqa}{\end{eqnarray}}
\newcommand{\eeqe}{\end{equation}}

\newcommand{\grad}{$^{\circ}$}

\begin{document}

\title{Large effect of the metal substrate on the magnetic anisotropy of Co on hexagonal Boron Nitride}

\author{Iker Gallardo$^{1,2}$, Andres Arnau$^{1,2,3}$, Fernando Delgado$^{2,4}$}

\address{$^{1}$ Centro de F\'{\i}sica de Materiales CFM/MPC (CSIC-UPV/EHU), 
Paseo Manuel de Lardiz\'abal 5, 20018 Donostia-San Sebasti\'an, Spain}
\address{$^{2}$ Donostia International Physics Center, 
Paseo Manuel de Lardiz\'abal 4, 20018 Donostia-San Sebasti\'an, Spain}
\address{$^{3}$ Departamento de F\'{\i}sica de Materiales, Facultad de Qu\'{\i}mica, 
Universidad del Pa\'{\i}s Vasco UPV/EHU,
Apartado 1072, 20080 Donostia-San Sebasti\'an, Spain}
\address{$^{4}$ Instituto de estudios avanzados IUDEA, Departamento de F\'{i}sica, Universidad de La Laguna, C/Astrof\'{i}sico Francisco S\'anchez, s/n. 38203, Tenerife, Spain}

\author{Romana Baltic$^{5}$, Aparajita Singha$^{6, 5, 7}$, Fabio Donati$^{6, 5, 7}$, Christian W{\"a}ckerlin$^{5, 8}$, Jan Dreiser $^{5, 9}$, Stefano Rusponi$^{5}$, Harald Brune$^{5}$}
\address{$^{5}$ Institute of Physics, Ecole Polytechnique F{\'e}d{\'e}rale de Lausanne, Station 3, CH-1015 Lausanne, Switzerland}
\address{$^{6}$ Center for Quantum Nanoscience, Institute for Basic Science (IBS), Seoul 03760, Republic of Korea}
\address{$^{7}$ Department of Physics, Ewha Womans University, Seoul 03760, Republic of Korea}
\address{$^{8}$ 
EMPA, Swiss Federal Laboratories for Materials Science and Technology, {\"U}berlandstrasse 129, 8600 D{\"u}bendorf, Switzerland}
\address{$^{9}$ Swiss Light Source, Paul Scherrer Institute, CH-5232 Villigen PSI, Switzerland}


\date{\today}

\begin{abstract}

We combine x-ray absorption spectroscopy (XAS), x-ray magnetic circular dichroism (XMCD) and x-ray magnetic linear dichroism (XMLD) data
with first principles density functional theory (DFT) calculations and a multiorbital many body Hamiltonian approach
to understand the electronic and magnetic properties of Co atoms adsorbed on h-BN/Ru(0001) and h-BN/Ir(111). The XAS line shape reveals, for both substrates, an electronic configuration close to $3d^8$, corresponding to a spin $S = 1$. Magnetic field dependent XMCD data show large (14 meV) out-of-plane anisotropy on h-BN/Ru(0001), while it is almost isotropic (tens of $\mu$eV) on h-BN/Ir(111). XMLD data together with both DFT calculations and the results of the multiorbital Hubbard model suggest that the dissimilar magnetic anisotropy originates from different Co adsorption sites, namely atop N on h-BN/Ru(0001) and 6-fold hollow on h-BN/Ir(111).
 
\end{abstract}


\maketitle



\section{Introduction} 
 
The stability of a magnetic moment against 
fluctuations critically depends on the magnetic anisotropy energy (MAE). 
At the microscopic scale, a large MAE may result from large spin-orbit coupling and suitable 
crystal field symmetry~\cite{Rau988,Donati_Rusponi_science_2016}. Therefore, although spin-orbit coupling in late 3d-transition metal atoms is relatively weak, 
under special circumstances the 3d adsorbates can show very large MAE values, as it is the case of Co single atoms and nanostructures on Pt single crystals~\cite{Gambardella2002,Gambardella1130}. However, large MAE do not guarantee stability, as it is the case of Co atoms on Pt(111), which have sub-nanosecond spin lifetime and do not show hysteresis even at sub-Kelvin temperatures~\cite{Mei08}. These observations highlight that 
spin scattering with conduction electrons and substrate phonons is detrimental to the stability of the magnetic moment of isolated impurities. Furthermore, atomic-scale magnetic moments can be quenched when interacting with the itinerant electrons on a nearby metal, leading to the formation of Kondo singlets~\cite{Hewson_book_1997}. Thus, large MAE and weak electronic and phonon interactions with the surroundings are key ingredients to achieve stable nanomagnets~\cite{Rau988,Donati_Rusponi_science_2016,Natterer_Yang_nature_2017}.
These considerations explain the interest in 3d and 4f atoms deposited on surfaces like MgO/Ag(100)~\cite{Rau988,PhysRevLett.115.237202,Pau17}, graphene~\cite{Don13,Don14,Bal16} or h-BN/Rh(111)~\cite{PhysRevLett.109.066101,jacobson2015,Muenks2017}. 
In these systems, the decoupling layer is introduced to reduce the hybridization of the adsorbate with the metallic substrate, thus protecting the magnetic moments from destabilizing scattering processes, while still allowing for 
large values of the MAE. For instance, in the case of h-BN/Rh(111) \cite{jacobson2015}, CoH magnetic impurities have been found to adsorb on atop N sites of the h-BN lattice and to show relatively large out-of-plane MAE close to 5 meV with spin quantum number S=1.

The MAE of single magnetic impurities adsorbed on surfaces can be measured mainly by two experimental techniques: x-ray magnetic circular dichroism (XMCD) and inelastic electron tunnelling spectroscopy (IETS). The interpretation of XMCD data relies on electronic multiplet calculations and the sum rules~\cite{Tho92,Car93} analysis with some fitting parameters required to determine the crystal field and spin-orbit coupling strength. 
The outcome are the spin and orbital angular moments together with the energy spectrum of ground and excited states. 
The other technique, i.~e., IETS~\cite{Heinrich_Gupta_science_2004}, yields the spin excitation energies of a single or a few magnetic coupled 
atoms with atomic resolution. Then, by using a phenomenological spin Hamiltonian compatible with the symmetry of the crystal environment, one can usually describe the magnetic anisotropy in terms of some characteristic anisotropy parameters~\cite{Rau988,PhysRevLett.115.237202,jacobson2015,Hirjibehedin_Lin_Science_2007,otte2008}.  

From the theoretical point of view, a recently developed approach~\cite{PhysRevB.92.174407,1367-2630-17-3-033020} consists 
%
%
in using first principles density functional theory (DFT) calculations to obtain the crystal field and, later on, with this information at hand, to construct a crystal field term in a many-body multiorbital Hamiltonian that also includes spin-orbit and electron correlation effects. Additionally, DFT calculations including spin-orbit coupling permit an estimation of the MAE but, in general, the so obtained values for magnetic adatom impurities adsorbed on surfaces are significantly lower than measured values due to the overestimation of orbital momentum quenching \cite{Blo10,PhysRevB.93.140101,PhysRevB.93.224428}.

In this work, we combine XAS, XMCD and XMLD measurements with a multiorbital Hamiltonian model and DFT calculations to disentangle the magnetic properties of Co atoms adsorbed on h-BN/Ir(111) and h-BN/Ru(0001) surfaces. h-BN is an insulating layer with a large band gap of about 6 eV, which can efficiently decouple the impurity from the metallic substrate. The substrate choice is dictated by the different binding energy between h-BN and the two single crystals~\cite{las08}, namely strong on Ru(0001) and weak on Ir(111). This is expected to promote different crystal fields (CF) and, thus, also different magnetic properties for the adsorbates. In agreement with these expectations, line shape and magnetic field dependence of XAS and XMCD measurements combined with multiplet calculations reveal a large out-of-plane anisotropy of about $14$ meV for Co on h-BN/Ru(0001), contrary to the low value of a few tens of $\mu$eV for Co on h-BN/Ir(111). 
Additional XMLD measurements at high and low external magnetic fields suggest that the dissimilar magnetic anisotropy originates from different Co adsorption sites, namely atop N and hollow on h-BN/Ru(0001) and h-BN/Ir(111), respectively. The spin and orbital magnetic moment of the Co atom, as well as the MAE, obtained from DFT calculations confirm this picture at a qualitative level. Moreover, the use of a many-body multiorbital Hamiltonian permits the achievement of a better quantitative agreement with experiment and a deeper understanding of two systems, in particular of their response to an external magnetic field.

\section{Methods}

\subsection{Sample Preparation}
The Ir(111) and Ru(0001) single crystals were prepared {\it in-situ} by repeated cycles of Ar$^+$ sputtering and annealing. Subsequently, {\it h}-BN was grown by chemical vapor deposition (CVD) using borazine (125 Langmuir at 1030~K). The reaction is self-limited to one monolayer since the catalytic dissociation of the precursor molecule requires bare metal areas. Co was deposited from a high purity rod (99.998 \%) using an $e$-beam evaporator in a background pressure of $\leq 3 \times 10^{-11}$~mbar. The Co coverage is expressed in monolayers (ML), where 1.0 ML is defined as one Co atom per h-BN unit cell.

\subsection{DFT calculations}

We have used the VASP package \cite{PhysRevB.59.1758,PhysRevB.54.11169,KRESSE199615} to perform \textit{ab initio} calculations based on DFT within the Projected Augmented Wave (PAW) method \cite{PhysRevB.47.558} and the Perdew-Burke-Ernzerhof (PBE) exchange-correlation functional\cite{PhysRevLett.77.3865}. The calculations have been done using a 4x4 h-BN supercell, which matches very well with a rotation of  13.9\grad of the Ir(111) and Ru(0001) surfaces. In this way, we avoid strain while having a computationally accessible unit cell. The effective vacuum region was larger than 7 \AA. A $\Gamma$-centred 3$\times$3$\times$1  \textit{k}-point sampling and an high energy ($600 ~eV$) cutoff were used. For the spin polarized calculations we also have included an on-site Coulomb interaction ($U=  4 ~eV$) to account for electron correlations in the Co 
\textit{3d} shell \cite{PhysRevB.57.1505}. 

Concerning the DFT calculations including spin orbit coupling (SOC), they have been done using the tetrahedron smearing method with Bl{\"o}chl correction \cite{PhysRevB.50.17953}.
DFT+SOC calculations can be used to extract information about the magnetic anisotropy of the system. This requires a very strict 
convergence, which was obtained by using an energy convergence threshold smaller than $10^{-6}$ eV. We achieve this convergence using $\Gamma$-centred 5$\times$5$\times$1, 7$\times$7$\times$1 and 9$\times$9$\times$1 \textit{k}-point samplings.

\section{Results and discussion} 

\subsection{XAS, XMCD and XMLD data \label{ExpData}}
\begin{figure}
\begin{center}
\includegraphics[width=0.75\textwidth]{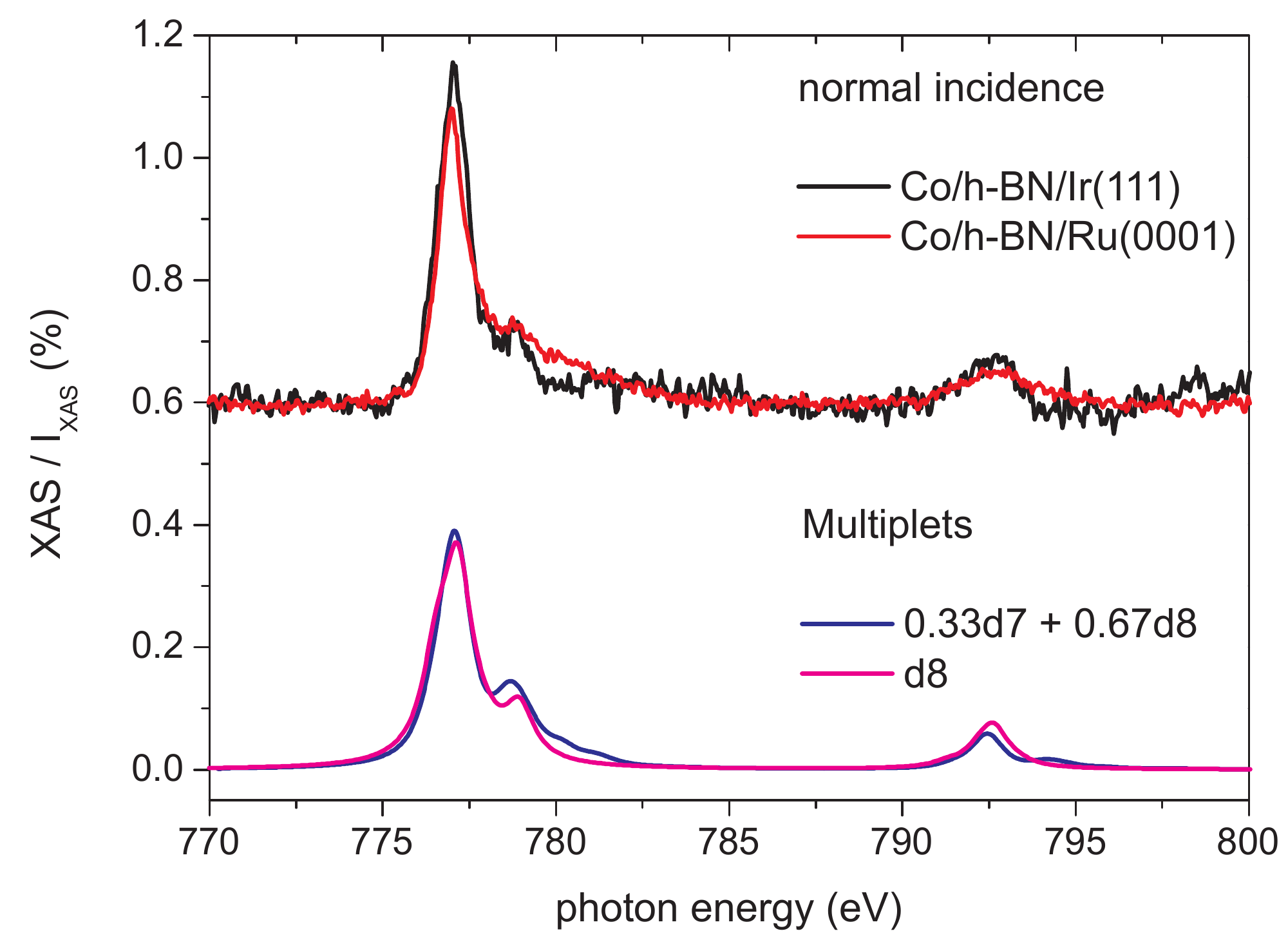}
\caption{(top) XAS spectra measured for Co on h-BN/Ir(111) and on h-BN/Ru(0001) ($\theta$ = 0$^\circ$, \textit{T} = 2.5 K, \textit{B} = 6.8 T, Co coverage $\Theta_{Co} = 0.005$ ML and $\Theta_{Co} = 0.008$ ML
on h-BN/Ir(111) and on h-BN/Ru(0001), respectively). Similar spectral features are observed on the two substrates. (bottom) Multiplet calculations with two different codes, namely CTM4XAS6 and multiX for mixed and pure \textit{d}-shell configuration, respectively. The electronic configuration of the Co atom is prevalently $3d^8$ corresponding to a spin quantum number of S = 1.}
\label{XAS}
\end{center}
\end{figure}

We investigate the electronic and magnetic properties of Co monomers created by low-temperature deposition on h-BN single layers grown by CVD on Ru(0001) and on Ir(111) (see methods for the sample preparation). Figure~\ref{XAS} compares the XAS spectra of isolated Co atoms deposited on the two surfaces with the spectra obtained from multiplet calculations with two different approaches. The experimental data show a fine multipeak structure, fingerprint of the Co electronic state. In particular we note that the $L_{\rm 3}$~edge splits in a main peak at 777.0 eV and in a small shoulder at 778.8 eV. This multipeak structure compared to the broad $L_{\rm 3}$ shape observed for Co atoms in bulk is the signature of an electronic state partially preserving an atomic like-character.
In order to disentangle the Co electronic state, we performed multiplet calculations using the CTM4XAS6~\cite{Sta10} and the multiX~\cite{uld12} code. In both cases, a CF with C$_{3v}$ symmetry is assumed, corresponding to Co adsorption on top of the N atom or in the hollow site. The CTM4XAS6 code describes the CF via Dq, Ds and Dt terms and allows for mixed configurations; the multiX code describes the CF via a point charge approach and it does not include partial charge transfer to the ligand, i.e., only pure 3$d^{\rm n}$ configurations are considered. Both calculations represent the data well and reveal a ground state configuration with mainly $3d^{\rm 8}$ character on both surfaces. This predisposition to acquire an extra electron in the 3d shell is frequently observed for Co atoms adsorbed on surfaces with low electron density at the Fermi level such as alkali metals~\cite{Gam02}, graphene~\cite{Don13,Don14,Bar16} or MgO~\cite{Rau988}.

\begin{figure}
\begin{center}
\includegraphics[width=0.99\textwidth]{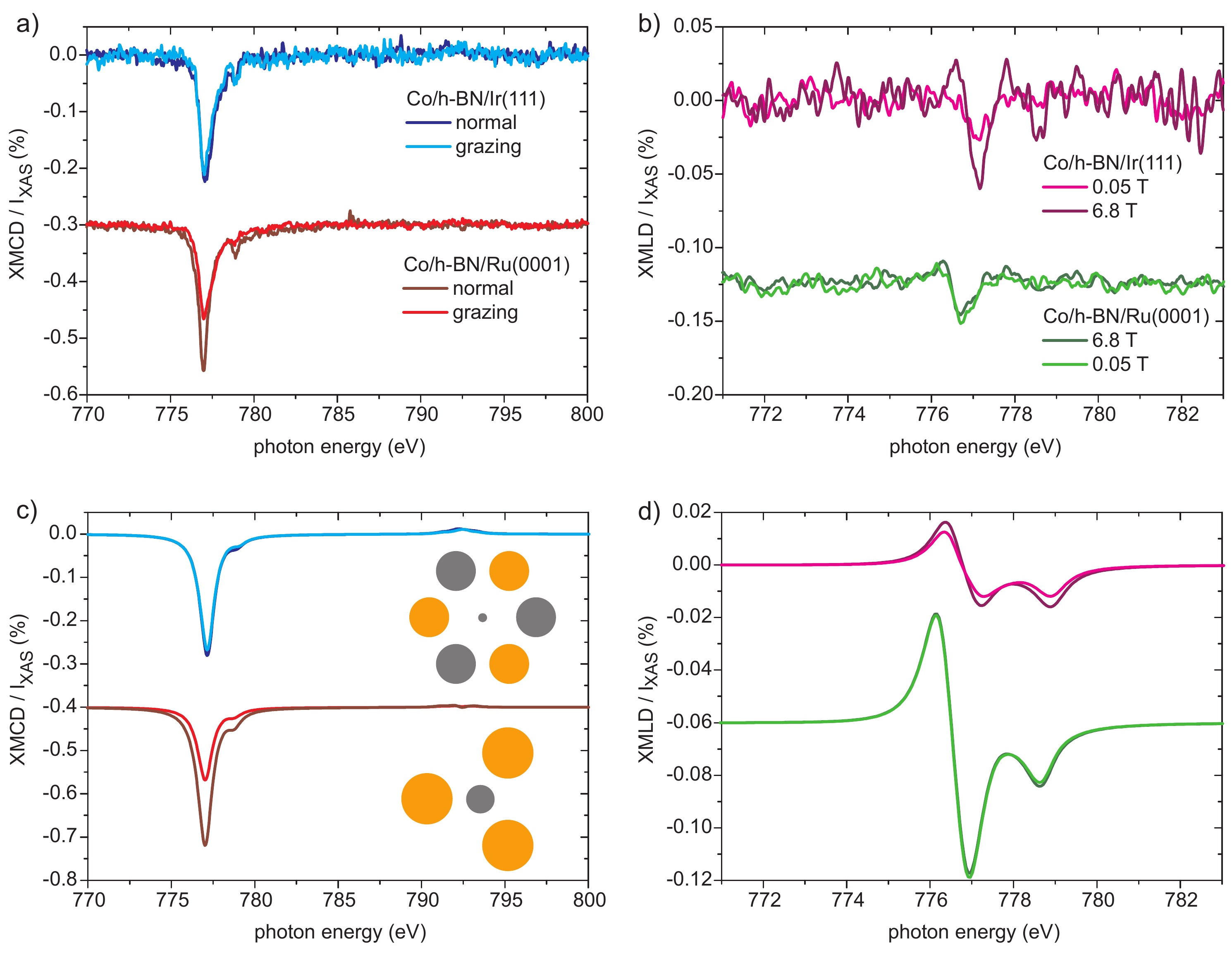}
\caption{(a) XMCD spectra measured for Co on h-BN/Ir(111) and on h-BN/Ru(0001) at normal and grazing incidence in an external field \textit{B} = 6.8 T. (b) XMLD spectra measured at grazing incidence for Co on h-BN/Ir(111) and on h-BN/Ru(0001) showing strong and weak field dependence, respectively. (c) XMCD and (d) XMLD spectra calculated with the multiX code by using the point charge distributions sketched in panel (c) for hollow and atop N Co adsorption sites. The area of the circles is proportional to the charge value, and the colour represents the sign of the charge (grey = positive, yellow = negative).}
\label{XLD}
\end{center}
\end{figure}

\begin{figure}
\begin{center}
\includegraphics[width=0.99\textwidth]{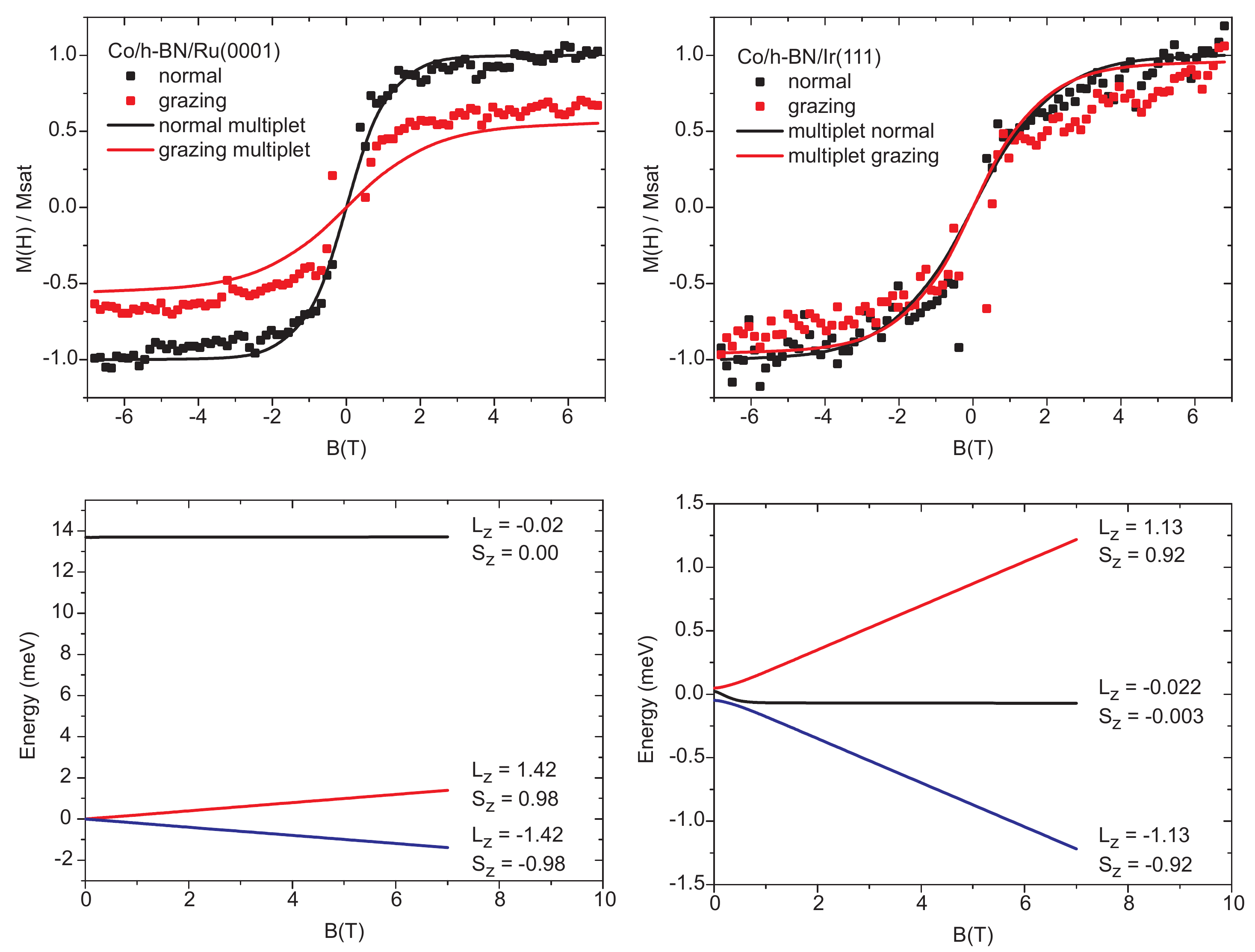}
\caption{ Upper panels: magnetization curves, (dot) experimentally acquired and (line) obtained from multiplet calculations, for Co on h-BN/Ru(0001) and h-BN/Ir(111) at both normal and grazing incidence .
Lower panels: Corresponding field splitting of the Co lower states obtained from multiX multiplet calculations. }
\label{MAG}
\end{center}
\end{figure}

Insights into the magnetic ground state and magnetic anisotropy are obtained by measuring XMCD and XMLD spectra at normal ($\theta$ = 0$^\circ$) and grazing ($\theta$ = 60$^\circ$) incidence as shown in Figure~\ref{XLD}. The quite similar XMCD shape and intensity observed at normal and grazing incidence for Co/h-BN/Ir(111) indicate negligible magnetic anisotropy. On the contrary, in the case of Co/h-BN/Ru(0001) we observe a strong angular dependence of the XMCD signal, larger at normal than at grazing incidence, indicating a strong out-of-plane magnetic anisotropy. These conclusions are also supported by the magnetic field dependent (independent) XMLD signal observed in Co/h-BN/Ir(111) (Co/h-BN/Ru(0001)). 
%
%
The microscopic origin of the magnetic anisotropy is the combined effect of the anisotropy in the atomic orbital moment dictated by the CF and the spin-orbit interaction ~\cite{Bru89,Laa98}. 
In solids, the orientation of the orbital moment is defined by the CF symmetry and strength. However, in an external magnetic field, S and L tend to align to the field itself; thus, the resulting configuration depends on the competition between CF and magnetic field. Linear dichroism is a measure of the charge density involved in perpendicular versus in-plane bonds forming between Co atoms and supporting substrate. 
%
%
Its field dependence, i.e. the XMLD, thus provides a measure of this competition. In systems with strong CF and magnetic anisotropy, the application of an external field can only marginally change the orientation of L and S, resulting in a field independent XMLD as observed in Co/h-BN/Ru(0001); the opposite behaviour is observed in low magnetic anisotropy systems. The angular dependence of the magnetization curves shown in Figure~\ref{MAG} fully confirm the low and high magnetic anisotropy scenario for Co/h-BN/Ir(111) and Co/h-BN/Ru(0001), respectively.

Comparison of the experimental data with multiplet calculations allows us to provide a more quantitative analysis. Given the mainly 3$d^{\rm 8}$ character of the Co atoms on both substrates, we focused on calculations with the multiX code in which a pure 3$d^{\rm 8}$ electronic state has been assumed. These calculations include the effect of the external magnetic field, finite temperature, incidence angle of x-rays, and crystal field environment of the magnetic atom. The CF is defined by effective point charges which position and intensity were chosen in order to simultaneously fit the shape and intensity of XAS, XMCD, and XMLD spectra as well as the shape of the magnetization curves at the two x-ray incidence angles. The point charge distributions that best reproduce the two systems are sketched in Figure~\ref{XLD}, the exact position and charge values are reported in the Appendix. They correspond to a hollow and top adsorption site for Co/h-BN/Ir(111) and Co/h-BN/Ru(0001), respectively. This is very similar to the behaviour of Co on graphene on both substrates
~\cite{Don14}. Orbital ($m_{L}$) and effective spin magnetic moment ($m_{S+D}$), given by the sum of the spin and dipolar term, projected onto the x-ray incidence direction are evaluated by applying the sum rules to the experimental and calculated spectra (see Table~\ref{table1}). The orbital moment is relatively large on both samples with values close to 1 $\mu_{B}$ suggesting an atomic like behaviour; however, it does not reach the close to free atom values observed for Co deposited on other decoupling layers such MgO~\cite{Rau988} or graphene on Ru(0001)~\cite{Don14}. In addition, $m_L$ shows a strong angular dependence for Co/h-BN/Ru(0001), with the largest value observed at normal incidence, while there is only a fractional reduction by moving from normal to grazing incidence for Co/h-BN/Ir(111). The different angular dependence of the orbital momentum observed in the two systems explains the high (negligible) magnetic anisotropy observed in Co/h-BN/Ru(0001) (Co/h-BN/Ir(111)), as also highlighted by the angular dependence of the magnetization curves in the two systems (see Figure~\ref{MAG}). We can quantify the strength of the MAE by the zero field splitting (ZFS), which is the energy difference between the ground and first excited state. The ZFS values, as deduced from the multiX calculations, are shown in Table~\ref{table1}. For Co/h-BN/Ru(0001) we find that the ground state consists of a doublet  ($S_{z} =  0.98, L_{z} = 1.42$) separated by 13.7 meV from the singlet ($S_{z} = 0, L_{z} = 0.02$). The ZFS of 13.7 meV obtained for Co/h-BN/Ru(0001) is quite large, similar to the values observed for Co on graphene supported by different metallic substrates~\cite{Don13,Don14}, and only lower than the ZFS of about 58 meV reported for Co/MgO/Ag(100)~\cite{Rau988}. On the contrary, for Co/h-BN/Ir(111) we find that the doublet ($S_{z} = 0.92, L_{z} = 1.13$) and the singlet ($S_{z} = 0.003, L_{z} = 0.022$) are very close in energy with a ZFS of only 70 $\mu$eV. We note that on both surfaces the lowest energy states are not pure $S_{z}=0,\pm 1$ and $L_{z}$ states, since they contain admixtures of different spins and orbital moments from electronic levels belonging to multiplets higher in energy.
 This admixture is also responsible for the energy splitting of the ground state doublet by 96.3 (6.6) $\mu$eV for Co/h-BN/Ir(111) (Co/h-BN/Ru(0001)), permitting 
 direct quantum tunnelling of the magnetization at zero magnetic field and thus leading to paramagnetic M(H) curves for both systems.

\begin{table*}[t!]
\begin{center}
\caption{Orbital ($m_{L}$) and effective spin ($m_{S+D}$) moments (in $\mu_{B}$), as well as their ratios, for normal (0$^\circ$) and grazing (60$^\circ$) incidence evaluated by applying the sum rules to the experimental (calculated) spectra assuming a hole number $n_{h} = 2$. The spin value and the ZFS as deduced from multiplet calculations are also reported.}
\vspace{0.5 cm}
\begin{tabular}{|l|c|c|c|c|c|c|c|c|c|c|c|c|}
	\hline
	& \multicolumn{3}{|c|}{Normal} & \multicolumn{3}{|c|}{Grazing}&\multicolumn{2}{|c|}{Multiplets} \\
	\hline
	& $m_{S+D}$ & $m_{L}$ & $m_{L}/m_{S+D}$ & $m_{S+D}$ & $m_{L}$ & $m_{L}/m_{S+D}$ & $S$ & $ZFS [meV] $ \\
	\hline 
	\hline
	h-BN/Ir(111) & $1.4 (2.01) $ & $0.8 (1.13) $ & $0.57 (0.56) $ & $1.2 (1.9) $ & $0.7 (1.10) $ & $0.58 (0.58)$ & $0.92$ & $0.07$ \\
	\hline
	h-BN/Ru(0001) & $1.4 (2.22)$ & $1.0 (1.42) $ & $0.71 (0.64) $ & $1 (1.25) $ & $0.6 (0.77) $ & $0.6 (0.62) $ & $0.98$ &  $13.7 $ \\
	\hline
\end{tabular}
\end{center}
\label{table1}
\end{table*}

\subsection{DFT calculations and construction of multiorbital model}

\subsubsection{DFT calculations} 

\begin{figure}
\begin{center}
\includegraphics[width=1.0\textwidth]{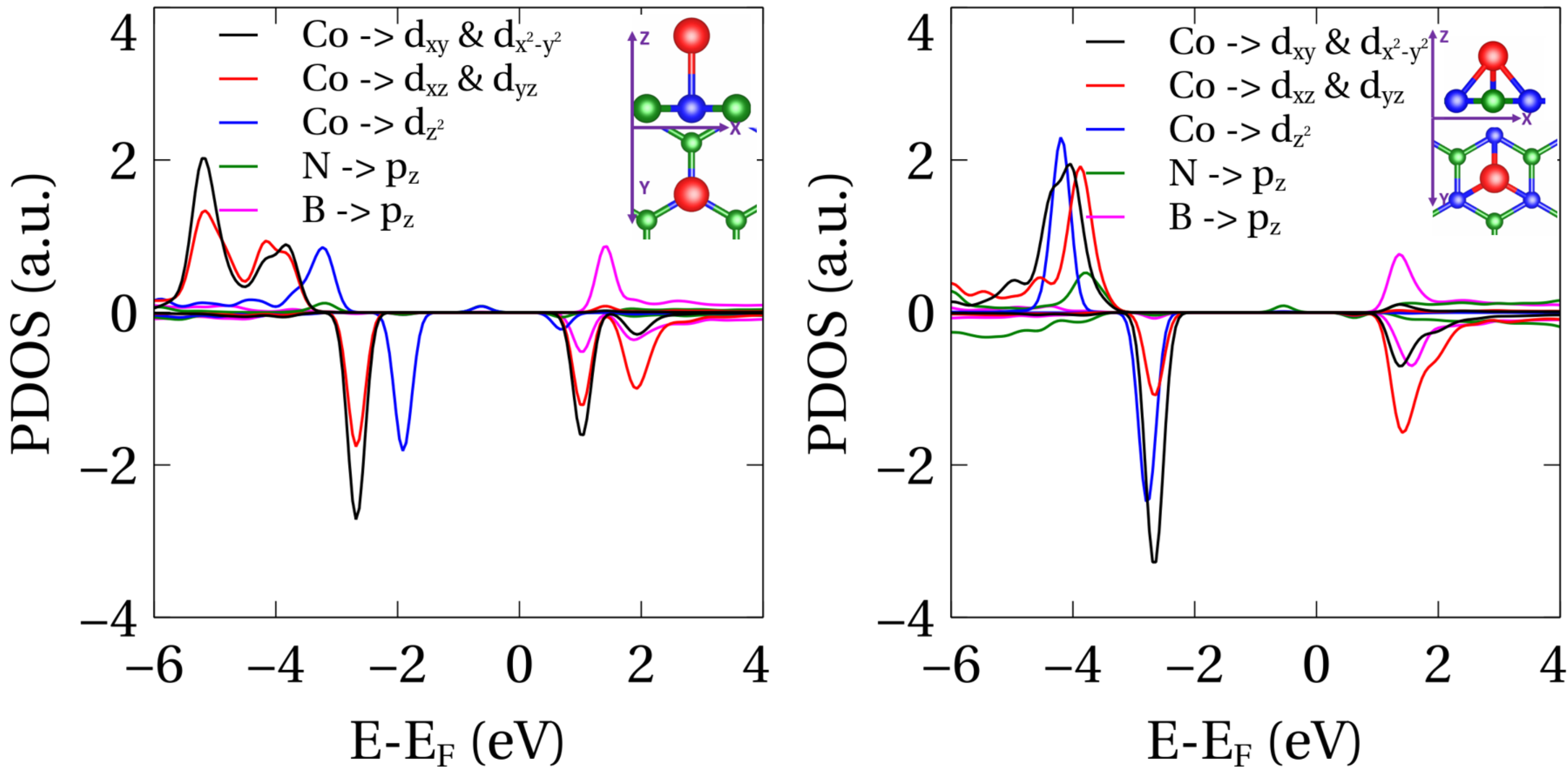}
\caption{\label{Configurations}Projected density of states (PDOS) onto 3d orbitals of Co and $p_{z}$ orbitals of first neighbour N and B atoms once for atop N (left) and once for hollow (right) adsorption sites, as  obtained from the spin-polarized DFT calculations. Positive values correspond to majority-spin and negative values to minority-spin.  The insets shows the adsorption site, where the red sphere represents the Co adatom, while the blue and green spheres represent N and B atoms, respectively. } 
\label{PDOS}
\end{center}
\end{figure}
\hfill \break
Spin polarized DFT+U calculations, including  vdW interaction, reveal an energetic preference for Co to adsorb on the 6-fold coordinated hollow sites  
of a free-standing h-BN layer, rather than atop N.  The corresponding adsorption energies 
differ by 
a few tenths of eV and both are lower than 1 eV, corresponding to weak chemisorption, where van der Waals forces are playing an important role. Indeed, a precise determination of the adsorption energies and the corresponding adsorption sites is not an easy task. The hollow adsorption site 
for Co on h-BN is in agreement with the result obtained by Yazyev and Pasquarello \cite{PhysRevB.82.045407}
who used 
similar computational methodologies.  Therefore, as h-BN is weakly bound on Ir(111), we assume the same Co hollow adsorption site also on h-BN/Ir(111).

However, the case of Co adsorption on h-BN/Ru(0001) corresponds to a more complex situation, which is at the limit of DFT reliability in giving the adsorption site. This is due to the fact that the hybridization between the h-BN layer and the Ru(0001) surface makes the h-BN/Ru(0001) surface more reactive than the h-BN/Ir(111) surface. A  similar situation 
was found 
for molecular adsorption on metal surfaces in the weak chemisorption regime, like CO on  Cu(111), where
the correct adsorption site is only obtained when hybrid functionals, like, BLYP, are used \cite{Wortmann2015}.
In case of h-BN/Ru(0001), additionally, the large size of the Moir\'e supercell formed by h-BN on Ru(0001)
makes the problem not tractable 
on the supercell of the real Moir\'e pattern.
Therefore, we follow a strategy similar to the one used to study the adsorption energy variation in different Moir\'e domains for h-BN/Ni(111) with a lattice matched model  \cite{PhysRevB.82.045407} but using a rotated 4x4 h-BN surface unit cell on Ru(0001) with different N-Ru registries, as described in the Methods section. The calculated variation in the adsorption energy of Co on h-BN/Ru(0001) for different registries is of the order of a few tenths of eV and, therefore, comparable to the difference in adsorption energy calculated by moving the Co atom from atop to hollow adsorption on h-BN.
Therefore, we prefer to use the information from XMCD and XMLD data to establish the adsorption site. 
Indeed, our multiorbital Hubbard model calculations described below are consistent with the observation of a large out-of-plane magnetic anisotropy for Co on h-BN/Ru(0001), when assuming atop N adsorption for the Co atom.

The calculated projected density of states (PDOS) onto 3\textit{d} states of Co  for the two adsorption sites on h-BN is shown in Figure   \ref{Configurations}. The energy integration of these PDOS curves onto 3\textit{d} states up to the Fermi level (\textit{E} = 0) gives values of about 7.8 electrons in the \textit{d} shell for the Co adatom (see Table \ref{tableDFT_CF}) in both cases, which approximately correspond to a spin S = 1 localized in the 3\textit{d} shell of the Co atom, in good agreement with experiment. In both cases, the appearance of relatively sharp peaks is consistent with a weak hybridization between the 3\textit{d} atom states and the h-BN states with  practically a full occupation of the majority spin 3\textit{d} states and of three minority spin states of Co. 
Both adsorption sites show weak hybridization of Co 3\textit{d} states with 2\textit{p} of first neighbour N and B atoms. This hybridization corresponds to $d_{z^{2}}$ of Co with $p_{z}$ of N atom for atop N adsorption site and $d_{yz}$ and $d_{xz}$ of Co with $p_{z}$ of N and B atoms for 6-fold coordinate hollow adsorption site.

Our DFT calculations including spin-orbit interaction permit us to estimate the MAE from the total energy difference of two self-consistent calculations corresponding to two different magnetization directions, in-plane and out-of-plane. The high precision required to obtain well converged MAE values and, thus, reliable results, restricts us in practice to consider only the Co atom adsorbed atop N and on a hollow site of a $4\times 4$ h-BN layer without the explicit inclusion of the metal surface atoms underneath. In this way, as described in the Methods section, we can do convergence checks with respect to both $k$-point sampling and energy cut-off in the plane wave expansion. This approximation is also justified by the fact that the magnetocrystalline anisotropy is  predominantly determined by the nearest neighbour atoms that define the crystal field. This extreme is further  confirmed by the multiorbital Hubbard model in the next section, where the crystal field term considers only four or six atoms for atop N or hollow adsorption respectively.  The well-known overestimation of the orbital momentum quenching by DFT calculations for this type of systems usually translates into an underestimation of the calculated MAE  and, thus, only the observed trend is reproduced: a significantly higher MAE value (1.5 meV) with out-of-plane anisotropy for Co adsorption on atop N sites compared to in-plane MAE for hollow site (0.4 meV). 
It is worth mentioning that the switching of the magnetization orientation from out-of-plane to in-plane when going from Co adsorption atop N site to Co adsorption on the hollow site, as well as the reduction of its magnitude by about a factor of three, is also found in our multiorbital Hubbard model. However, the corresponding ZFS values are an order of magnitude larger   and thereby in good agreement with XMCD data for the case of Co adsorption on atop N site (Co/h-BN/Ru(0001)) but not in the case of Co adsorption on the hollow site (Co/h-BN/Ir(0001)). 

\begin{table*}[t!]
\begin{center}
\caption{Summary of spin-moment ($m_{S}$), orbital moment ($m_{L}$) and zero-field splitting (ZFS) of Co atoms on an h-BN layer obtained from DFT calculations including the spin-orbit interaction. Here HA and EA stand for hard axis and easy axis, respectively.
}
\vspace{0.5 cm}
\begin{tabular}{|c|c||c|c|c|}
\hline
Adatom & Adsorption-site & $m_{S} \left(\mu_{B}\right)$ & $m_{L}$ $\left(\mu_{B}\right)$& ZFS (meV) \\
\hline \hline 
\multirow{2}{*}{Co} & atop N & $2.31$ & $0.29$ & $1.5$ (HA) \\
\cline{2-5}
& hollow & $2.20$ & $0.184$ & $0.43$ (EA) \\
\hline
\end{tabular}
\label{table2}
\end{center}
\end{table*}


\subsubsection{Spin excitation spectra from multiorbital Hubbard model \label{SecHModel}}
\hfill \break
In order to calculate the low energy excitations of the Co atoms adsorbed on the surface  we use a multiorbital Hubbard model. The purpose then is to build a many-body Hamiltonian derived from DFT calculations that accounts for the strong correlations in the system. This approach is adequate to describe spin excitations when charge redistribution and lattice deformation are negligible. 
Considering the information extracted from the fitting of the XMCD profile to the multiplet results of section \ref{ExpData}, we concentrate on describing the correlations of the partially-filled Co $d$-shell electrons ($d^8$-electronic configuration). 
We describe the interacting $N_e=8$ electrons at the  Co $d$-shell
by the Hamiltonian 
\begin{equation} \label{eq:2.100}
H = H_{\rm Coul} + H_{\rm CF} + \lambda_{\rm SO}H_{\rm SO} + H_{\rm Zeem}.
\label{Hmanyb}
\end{equation}
$H_{\rm Coul}$ refers to the electron-electron Coulomb interaction, $H_{\rm CF}$ is the crystal  
field contribution while $H_{\rm SO}$ and $H_{\rm Zeem}$ represents the spin-orbit and Zeeman interaction, respectively. The dimensionless $\lambda_{\rm SO}\in [0-1]$ coupling constant  is introduced to switch on and off the spin-orbit coupling when needed for the analysis of the results.
Electron-electron interaction is accounted for only in the Co  $3d$-orbitals, while the strength of the Coulomb interaction is determined by the Slater integrals $F^n(3d)$~\cite{slater1974quantum} or, similarly, in terms of an effective Hubbard repulsion $U$~\cite{1367-2630-17-3-033020}.  
Here, we take  
the values $F^0=3.41$ eV, $F^2=0.36$ eV and $F^4=0.13$ eV.\footnote{Our results do not show a significant variation with the Slater integrals as long as the Hund's rules are satisfied.}
The spin-orbit and the Zeeman interaction in the presence of an external magnetic field are evaluated following Refs.~\cite{1367-2630-17-3-033020,PhysRevB.92.174407}. 

The crystal field Hamiltonian is estimated using a point-charge model including only the first N and B neighbouring atoms. Both, charges and positions extracted from the DFT calculations are used for the computation, while the atomic-cloud dependent expectation values $\langle r^2\rangle$ and $\langle r^4\rangle$ are taken from the atomic values~\cite{Abragam_Bleaney_book_1970}. The single particle spin-orbit coupling constant ($\lambda_{\rm SO}=1$) is taken as the atomic value $\xi_{\rm Co}=65.5$ meV~ \cite{Abragam_Bleaney_book_1970}. Although it is well known that these point charge models do not provide a good quantitative description, especially when covalent bonds are present, they yield the right qualitative behaviour and they correctly reproduce the symmetry of the environment. Quantitative aspects will be discussed hereafter.

\begin{figure}
\begin{center}
\includegraphics[width=0.9\textwidth]{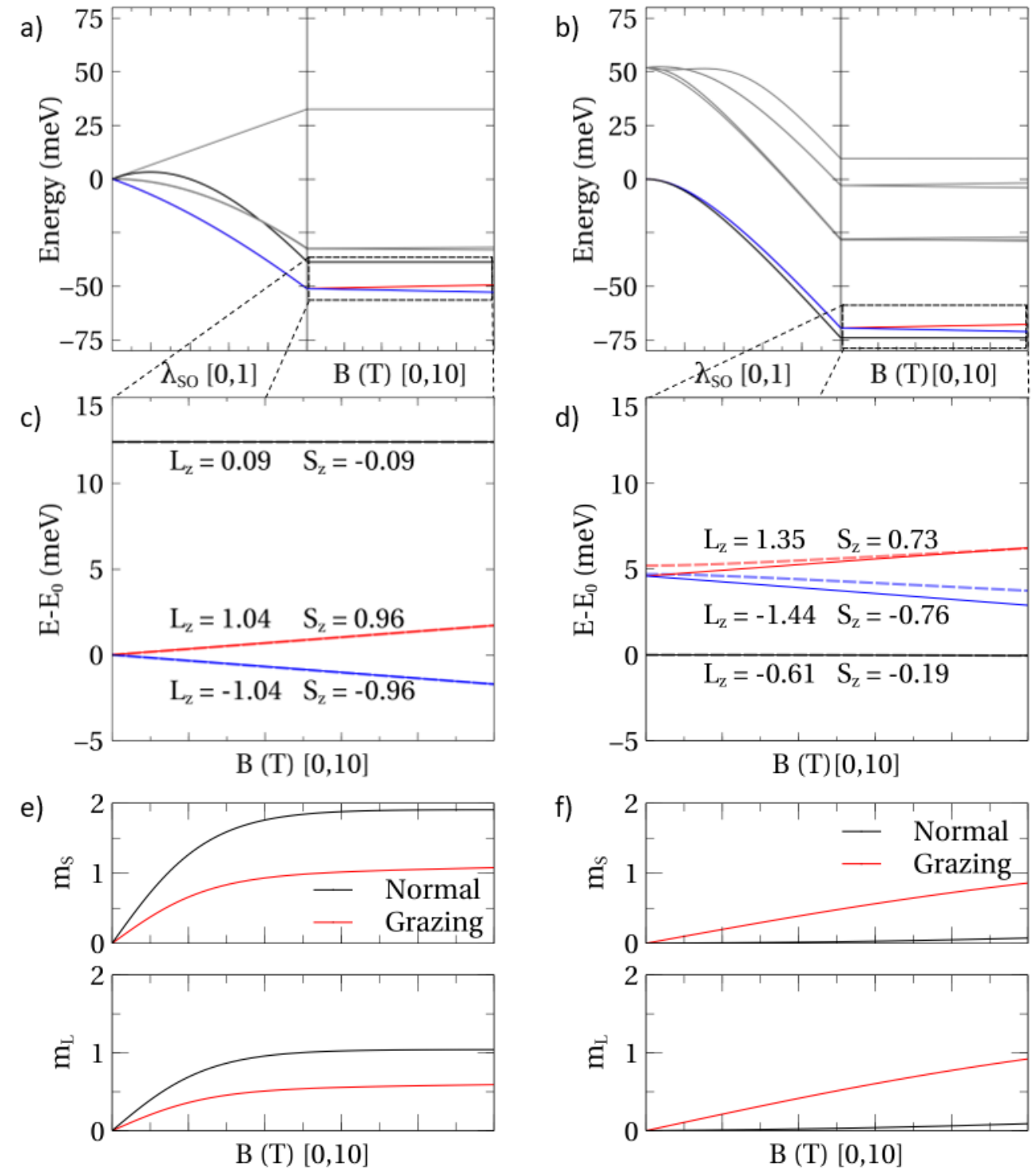}
\caption{\label{Co_multiplet} 
(a) and (b) 
Multiplet energy spectra $E_n$ versus spin-orbit coupling strength $\lambda_{SO}$ and magnetic field along the out-of-plane direction for Co adsorbed atop N and on hollow sites of h-BN, respectively. The solid blue, red and black curves correspond to the lowest  three energy states, while the light grey lines correspond to higher energy states (not considered in our discussion). 
(c) and (d) 
Zooms of the low energy sector for the three lower energy states corresponding to the effective $S=1$ anisotropic spin, where the labels correspond to the spin $S_{z}$ and orbital $L_{z}$ moments at $B=6.8$ T. The superposed dashed lines correspond to the solutions of the spin Hamiltonian (\ref{HSpin}).
(e) and (f)  
Average spin $m_{S}$  (left) and orbital $m_{L}$ (right) angular momenta in the direction of the applied  $B$ field for out of plane and grazing directions at $T=2.5 $ K.
}
\end{center}
\end{figure}

Figure \ref{Co_multiplet} shows the results of our multiplet calculation with the effective multiorbital model for Co on pristine h-BN on two different adsorption sites: atop N (left panels) and on a hollow site (right panels). In both cases, the total angular momentum and spin quantum numbers are $L=3$ and $S=1$, as expected 
for an atomic $3d^8$ configuration on the basis of Hund rules, but $|L_{z}|=1$ for the ground state. The crystal field splits the ground state multiplet, with degeneracy $(2L+1)(2S+1)=21$, in a different way on the two sites. For the atop N site, the ground state multiplet is an orbital doublet, with total degeneracy 6.  By contrast, the hollow site leads to an orbital singlet ground state. As a result, the spin-orbit coupling induces a qualitatively different behaviour in both systems. 

As observed from the magnetic field dependence, for the atop N position, the lowest doublet is formed by the $S_z=\pm 1$ states, while the excited state corresponds to $S_z\approx 0$, indicating an out-of-plane easy axis system, in good agreement with the data shown in Figure 3. This situation is reversed for the hollow absorption site, where an out-of-plane hard axis is found.  The change in the preferential magnetization axis between the two adsorption sites is also corroborated by 
the average magnetization along the applied magnetic field, see Figure~\ref{Co_multiplet}(c) and (d). 
Interestingly, the orbital moment is drastically affected by the adsorption site: the atop N  site leads to a significantly larger orbital moment, a situation reported before in similar systems with very high symmetry~\cite{Rau988,Etz15}. The origin of this unquenched orbital moment is the  large and almost perfectly axial crystal field created by the underlying N atom. On the contrary, the hollow site corresponds to a much lower point symmetry in which both 
the spin and orbital components along the field direction are quenched compared to the atop N site. 

The calculated average spin and orbital moments for the atop N site are in qualitative  agreement with the experimental results for Co/h-BN/Ru(0001),
 showing a saturation for fields around $3$ T, with a ratio between orbital ($m_{L}$) and spin moment ($m_{S}$), at both normal and grazing, of about 0.54, close to the experimetal values of 0.57 and 0.71 from Table~\ref{table1}. In contrast, the XMCD data for Co/h-BN/Ir(111) give a magnetic anisotropy much smaller than the one calculated for Co on the h-BN hollow site. Below we discuss some possible reasons.
There are two additional key considerations to be done when comparing the experimental results with the multiorbital Hubbard model.  
The presence of the underlying metal surface partially reduces the symmetry of the Co environment. Furthermore, the adsorption on different regions of the Moir\'e pattern can also induce important changes in the crystal field felt by the Co impurities. 
For adsorption on the atop N site, since the dominant source of the crystal field is the underlying N atom, a corrugation of the h-BN layer is not expected to lead to qualitative changes, with only small changes of the MAE between different Co impurities on the h-BN/Ru(0001) surface, assuming the same N-Ru registry. 
In contrast, on the hollow site the contributions of the six first neighbours are of similar magnitude which, together with the symmetry and the opposite charges on the B and N atoms, tends to cancel the electrostatic potential on the Co site. 
This situation can also be anticipated from the smaller energy splittings between the \textit{d}-orbital peaks in the PDOS, see Figure ~\ref{PDOS}.
Hence, a small corrugation, charge transfer, or lattice strain can induce important qualitative changes for atoms adsorbed on hollow positions, which might partially explain the apparent discrepancy for the Co hollow site adsorption on h-BN/Ir(111).   

A similar argument can also be applied to explain the deviations of the spin and orbital moments for Co/h-BN/Ir(111), especially for normal incidence. The moir\'e induced deviations from a perfect $C_{3v}$-symmetry yield a finite quantum tunnelling splitting between the states with $S_z=\pm 1$ and, thus, a zero expectation value of $\langle S_a\rangle$  and $\langle L_a\rangle$  ($a=x,y,z$). The higher the deviations, the larger the splitting.  The external field can partially recover the saturation values only if the induced Zeeman splitting is larger than the quantum spin tunnelling splitting.  

Considering the approximations done and the simplicity of the model used in this section, the agreement with the experiment, see Table~\ref{table1} and Figure \ref{MAG}, is remarkable for the atop N site. Notice that despite the similarities in the multiplet fitting of Sec.~\ref{ExpData} and the multiorbital model employed here, in the latter case we are not using any fitting parameters, and the only free parameters are the Slater integrals that play a minor role, while all other variables entering into the model are taken as their atomic counterparts.

\subsubsection{Spin Hamiltonian extracted from the multiorbital model}
\hfill \break
Excitation spectra of diluted magnetic centres in a paramagnet are often described in terms of spin Hamiltonians~\cite{Abragam_Bleaney_book_1970} , which depend only on the spin degrees of freedom. In the case of a spin $S=1$ system,  the spin Hamiltonian can be written as  
\beqa
H_S=D \hat S_{z'}^2+E\left(\hat S_{x'}^2-\hat S_{y'}^2\right)+\mu_B  \vec B\cdot g \cdot \vec S,
\label{HSpin}
\eeqa
where $\hat S_a$ is the a-component of the spin operator and $D$ and $E$ are the axial and transverse magnetic anisotropy parameters [($|D|=$ZFS for a spin $S=1$)].
At finite external field $\vec B$, the spectrum changes due to the induced Zeeman splitting, with a response characterized by the Land\'e g-factor tensor. It should be pointed out that, although we may not expect the transverse term in (\ref{HSpin}) from the $C_{3v}$-symmetry of the two adsorption sites considered here, this term can appear due to any symmetry breaking, for instance, by the Ru(0001) underneath or by any surface stress. This situation has also been found for Co/h-BN/Rh(111)~\cite{jacobson2015}. 

The usual approach is to estimate the parameters in (\ref{HSpin}) from fitting to experiments. Here, however, we take a radically different approach. We construct such a Hamiltonian from the many-body Hubbard Hamiltonian. If we denote by $|E_n\rangle$ and $E_n$ the eigenvectors and eigenvalues of the many-body Hamiltonian (\ref{Hmanyb}), and taking advantage that $\langle E_n|\hat S^2|E_n'\rangle \approx S(S+1)\delta_{nn'}$, we can construct $H_S$ by projecting the $(2S+1)$ low energy states into the bases of eigenstates of $\hat S^2,\hat S_{z'}$, i.e.,
$H_S=\sum_{n=1,\dots,2S+1}E_n \hat {\cal P}_{S}|E_n\rangle \langle E_n|\hat {\cal P}_S$,
where $ \hat{\cal P}_{S}=\sum_{m_{z'}}|S,m_{z'}\rangle\langle S,m_{z'}|$. We notice that the quantization axis $z'$ may be different from the $z$-axis taken in the DFT calculations, see inset Figure \ref{Configurations}, in which case some rotations may be needed to recover the simple form (\ref{HSpin}). Table \ref{tableHS} summarizes the parameters found from our analysis. In agreement with the previous results, the Co atop N can be described as an easy axis system with the $\hat z'$ direction out-of-plane. By contrast, the Co on hollow site corresponds to a hard axis. 
The energy spectra of $H_S$ are also depicted in Figure \ref{Co_multiplet} a) and b) with dashed lines.
The accordance between these results and those of the multiorbital Hubbard model indicates the robustness of our assumptions in deriving $H_S$.

\begin{table}[]
\begin{center}
\caption{Parameters of the spin Hamiltonian $H_S$ in equation (\ref{HSpin}) as extracted from the multiorbital Hubbard model. \label{tableHS}}
\begin{tabular}{|l|c|c|c|c|c|}
\hline
& $D$ (meV) & $E$ (meV) & $g_{xx}$ & $g_{yy}$ & $g_{zz}$ \\ \hline
Atop N & -12.41 & 0.008 & 1.72 & 1.67 & 2.96 \\ \hline
Hollow & +3.34 & 0.50 & 3.15 & 3.01 & 2.15 \\ \hline
\end{tabular}
\end{center}
\end{table}

\section{Conclusions} 

XAS, XMLD and XMCD measurements reveal large out-of-plane magnetic anisotropy for Co individual atoms adsorbed on h-BN/Ru(0001), while Co atoms on h-BN/Ir(111) have basically no anisotropy. This surprising finding is explained using a combination of first principles DFT calculations and a multiorbital Hubbard Hamiltonian.
The most important result of spin polarized DFT calculations is the determination of the spin $S=1$ of Co in agreement with the XAS data.
In addition, the adsorption sites for Co on h-BN/Ru(0001) and on h-BN/Ir(111) are identified as atop N and hollow sites, respectively, via the construction of a crystal field Hamiltonian that reproduces correctly the essential trends in magnetic anisotropy.
The deduced Co adsorption sites for the two surfaces are further supported by comparison with the results of the multiorbital Hubbard model. This also leads to a large easy axis MAE for the atop N position, and a smaller in-plane MAE with hard axis for the hollow site. Our results illustrate how two different underlying metal surfaces can lead to dramatically different magnetic anisotropy energies. These differences are mainly attributed to the different crystal field of the different adsorption sites, while the retained orbital moments and \textit{d}-shell filling are similar.

\section{Acknowledgments}

I. G. and A. A. acknowledge financial support by the  Spanish Ministerio de Economia y Competitividad (MINECO Grant No. FIS2016-75862-P). The calculations were performed using the Computer Center at the  Donostia International Physics Center (DIPC).
FD acknowledges funding from the Ministerio de Ciencia e Innovaci\'on, grant MAT2015-66888-C3-2-R and FEDER funds;  Basque Government, grant IT986-16, and Canary Islands program {\it Viera y Clavijo}  (Ref. 2017/0000231).
We gratefully acknowledge funding by the Swiss National Science Foundation (SNSF) through Grants No.200021\_146715/1 (R.B.), No. 200020\_157081/1 (A.S.), and No. PZ00P2\_142474 (C.W. and J.D.)
\newpage

\appendix
\setcounter{section}{1}
\section*{Appendix}

\subsection{X-ray Absorption Spectroscopy}
The X-ray absorption measurements were performed at the EPFL/PSI X-Treme beamline at the Swiss Light Source, Paul Scherrer Institut, Villigen, Switzerland~\cite{pia12}. The experiments were carried out in the total electron yeld (TEY) mode at 2.5~K, for circularly ($\sigma^+$, $\sigma^-$) and linearly ($\sigma^h$, $\sigma^v$) polarized x-rays, with the magnetic field applied parallel to the incident X-ray beam. The XAS corresponds to ($\sigma^+ + \sigma^-$), while the XMCD and XMLD correspond to ($\sigma^+ - \sigma^-$) and ($\sigma^v - \sigma^h$), respectively. The TEY mode enables the high sensitivity required by the extremely low concentration of magnetic elements at the surface. The XAS spectra were acquired with the magnetic field collinear with the photon beam at normal ($\theta = 0^{\circ}$) and grazing incidence ($\theta = 60^{\circ}$). To take into account the different surface areas illuminated by the X-ray beam at both sample orientations, the XAS spectra were normalized to the intensity at the pre-edge (772~eV). Prior to deposition of Co, background spectra on the given substrate have been recorded and then subtracted from the Co XAS spectra to eliminate the substrate contribution. 

\subsection{Multiplet simulations with the CTM4XAS6 code}
The CTM4XAS6 code~\cite{Sta10} is based on an atomic multiplet model which takes into account electron-electron interaction, charge transfer to the ligand through configuration interaction, crystal field potential and spin-orbit coupling.  Rescaled Slater-Condon integrals account for \textit{p} and \textit{d} electron-electron interaction.  Charge transfer to the surface is enabled via hopping terms reflecting the symmetry of
the atomic environment. Both the adsorption on top and hollow h-BN sites correspond to a CF with $C_{3v}$ symmetry. However, we model the crystal field potential in a $C_{\infty v}$ symmetry. This uniaxial crystal field lifts the degeneracy of the $d$-orbitals producing an $a_1$ singlet ($d_{\rm z^2}$) and two doublets $e_1$ ($d_{\rm xz}, d_{\rm yz}$) and $e_2$ ($d_{\rm x^2-y^2},d_{\rm xy}$). The $C_{3v}$ transverse term preserves the two-fold degeneracy of the $e_1$ and $e_2$ doublets 
%
%
; thus, it is expected to have a small effect on the total spin and orbital moments and it is hence neglected. XAS and XMCD spectra were best reproduced using a mixed $d^7 + d^8 L$ configuration with crystal field terms $10Dq = 0.0$~eV, $Ds = 0.36$~eV and $Dt = 0.2$~eV. Configuration interaction is allowed respecting the degeneracy of the $d$-orbitals, therefore we used the hopping integrals $t(a_{1})~=$~0.8 eV, $t(e_{1})~=$~0.4 eV and $t(e_{2})~=$~0.35 eV. The charge transfer energy $\Delta $ and the core-hole interaction $U_{p-d}-U_{d-d}$ were set to $-2.0$ eV and 1.1 eV, respectively. Default values for the Slater-Condon integrals have been used, corresponding to a reduction to 80\% of the Hartree-Fock values. Transition amplitudes for $L_{2}$ and $L_{3}$ edges were calculated in a dipolar approximation and broadened with Lorentzian functions of FWHM = 0.9 and 0.4 eV, respectively, to reproduce the experimental spectra. Further gaussian broadening of 0.4 eV FWHM was introduced to include the finite experimental energy resolution.

\subsection{Multiplet simulations with the multiX code}
The multiX code~\cite{uld12} uses an effective point charge approach for the description of the crystal field generated by the interaction of the Co atom with the surrounding substrate atoms. The position and intensity of the effective charges were chosen in order to simultaneously fit the shape and intensity of XAS, XMCD, and XMLD spectra as well as the magnetization curves, which display the magnetic field dependence of the XMCD peak intensity. Table~\ref{tableCF} summarizes the spatial distribution and the values used for the effective charges on the two surfaces. The values of the spin-orbit coupling and Coulomb interactions were scaled to 95 \% and 80 \% of the Hartree-Fock values, respectively. The experimental line broadening due to the finite lifetime of the core-hole state was modelled by convolution with a Gaussian of FWHM = 0.5~eV.

\begin{table*}[h!]
\caption{Point charge CF scheme employed in multiplet calculations with the multiX code.}
\vspace{0.5 cm}
    \centering
  \label{tab3}
\begin{tabular}{|c|c|c|c|c|c|c|c|c|}
     
     \hline
		\multicolumn{4} {|c|} {h-BN/Ir(111)} & \hspace{5mm} & \multicolumn{4}{c|} {h-BN/Ru(0001)}\\
		\hline
		\hline
     $x$ (\AA)  &  $y$ (\AA)  &  $z$ (\AA)  & $q$ ($e$)  & & $x$ (\AA)  &  $y$ (\AA)  &  $z$ (\AA)  &  $q$ ($e$)  \\
	
 \hline
$0$		  & $0$     & 	$-1.70$ 	&	$0.10$  & & $0$	    & $0$     &	$-1.60$ &	$1.0$\\
$1.50$	& $0$     & 	$-1.70$ 	&	$-1.40$ & & $1.50$	& $0$     &	$-1.60$ &	$-1.80$\\
$-0.75$	& $1.30$  & 	$-1.70$ 	&	$-1.40$ & & $-0.75$ & $1.30$  &	$-1.60$ &	$-1.80$\\
$-0.75$	& $-1.30$ & 	$-1.70$ 	&	$-1.40$ & & $-0.75$ & $-1.30$ &	$-1.60$ &	$-1.80$\\
$-1.50$	& $0$     & 	$-1.70$ 	&	$1.40$  & & $-$	    & $-$     &	$-$	    &	$-$\\
$0.75$	& $1.30$  & 	$-1.70$ 	&	$1.40$  & & $-$	    & $-$     & $-$     &	$-$\\
$0.75$	& $-1.30$ & 	$-1.70$ 	&	$1.40$  & & $-$	    & $-$     &	$-$     &	$-$\\
  \hline
    
\end{tabular}
\label{tableCF}
\end{table*}

\subsection{DFT values of Co electronic charge and magnetic moment}

\begin{table*}[h!]
\caption{Populations and spin polarization of Co orbitals from spin-polarized DFT calculations.}
\vspace{0.5 cm}
\centering
\begin{tabular}{|c c c c|}
\hline 
 & Total & 4s  & 3d \\
 \hline
Total charge & 8.77 & 0.98 & 7.79 \\
\hline
Spin up & 5.80 & 0.87 & 4.93 \\
\hline
Spin down & 2.97 & 0.11 & 2.86 \\
\hline
Polarization & 2.83 & 0.76 & 2.07 \\
\hline
\end{tabular}
\label{tableDFT_CF}
\end{table*}

\subsection{Multiorbital Hubbard model}
 
We describe the crystal field of the Co adatom using an effective point charge model, whose parameters were extracted from spin polarized DFT calculations. The charges and their positions are summarized in Table~\ref{tableDFT_CF}.

\begin{table*}[h!]
\caption{Position and charges of the point-charges used to calculate the CF contribution in the multiorbital Hubbard model.}
\vspace{0.5 cm}
    \centering
  \label{tab4}
\begin{tabular}{|c|c|c|c|c|c|c|c|c|}
     
     \hline
		\multicolumn{4} {|c|} {Hollow site} & \hspace{5mm} & \multicolumn{4}{c|} {Atop N site}\\
		\hline
		\hline
     $x$ (\AA)  &  $y$ (\AA)  &  $z$ (\AA)  & $q$ ($e$)  & & $x$ (\AA)  &  $y$ (\AA)  &  $z$ (\AA)  &  $q$ ($e$)  \\
	
 \hline
$1.23$		  & $0.71$     & 	$-1.66$ 	&	$-0.50$  & & $0.00$	    & $0.00$     &	$-1.90$ &	$-1.50$\\
$0.00$	& $-1.42$     & 	$-1.66$ 	&	$-0.50$ & & $1.51$	& $0.00$     &	$-1.90$ &	$0.50$\\
$-1.23$	& $0.71$  & 	$-1.66$ 	&	$-0.50$ & & $-0.75$ & $1.30$  &	$-1.90$ &	$0.50$\\
$0.00$	& $1.42$ & 	$-1.68$ 	&	$0.50$ & & $-0.75$ & $-1.30$ &	$-1.90$ &	$0.50$\\
$1.23$	& $-0.71$     & 	$-1.68$ 	&	$0.50$  & & $-$	    & $-$     &	$-$	    &	$-$\\
$-1.23$	& $-0.71$  & 	$-1.68$ 	&	$0.50$  & & $-$	    & $-$     & $-$     &	$-$\\
  \hline   
\end{tabular}
\label{tableDFT_CF}
\end{table*}




\newpage

\providecommand{\newblock}{}

\end{document}